# Resource-efficient Transmission of Vehicular Sensor Data Using Context-aware Communication

**Benjamin Sliwa[1], Thomas Liebig[2], Robert Falkenberg[1], Johannes Pillmann[1] and Christian Wietfeld[1]**
[1]Communication Networks Institute, [2]Department of Computer Science VIII
TU Dortmund University, 44227 Dortmund, Germany
e-mail: {Benjamin.Sliwa, Thomas.Liebig, Robert.Falkenberg, Johannes.Pillmann, Christian.Wietfeld}@tu-dortmund.de

*Abstract*—Upcoming Intelligent Traffic Control Systems (ITSCs) will base their optimization processes on crowdsensing data obtained for cars that are used as mobile sensor nodes. In conclusion, public cellular networks will be confronted with massive increases in Machine-Type Communication (MTC) and will require efficient communication schemes to minimize the interference of Internet of Things (IoT) data traffic with human communication. In this demonstration, we present an Open Source framework for context-aware transmission of vehicular sensor data that exploits knowledge about the characteristics of the transmission channel for leveraging connectivity hotspots, where data transmissions can be performed with a high grade if resource efficiency. At the conference, we will present the measurement application for acquisition and live-visualization of the required network quality indicators and show how the transmission scheme performs in real-world vehicular scenarios based on measurement data obtained from field experiments.

## I. INTRODUCTION

With massive deployment of cars used as mobile sensor nodes proving information for crowdsensing-based cloud services, the efficiency of the communication becomes a major challenge as this kind of MTC is usually performed in public cellular networks, leading to a competition of IoT devices and human participants among the available network resources. Usually, the obtained data is transmitted in a periodical way without considering the properties of the radio channel. Consequently, many transmissions happen during low quality periods of the channel and waste spectrum- and energy resources. To overcome this issue, the *anticipatory* communication scheme has been proposed [1]. Instead of immediately transmitting the measured information, data is aggregated locally if the current channel quality is low and transmitted when the vehicle experiences a *connectivity hotspots*, where transmissions can be performed in a very resource-efficient way. While this approach introduces another delay into the processing chain due to the local buffering, it satisfies the requirements of most crowdsensing-based services (e.g. traffic control, road quality estimation and weather forecast), as these work well with data that is up-to-date in the dimension of several minutes.

In recent work [2], we proposed a context-aware data transmission scheme as an extension to Channel-aware Transmission (CAT) [3] that uses machine learning based throughput prediction in order to optimize the transmission time and avoid transmissions during low channel quality periods. In this paper, the software framework for measuring and evaluation of the context-aware transmission schemes is demonstrated. The framework itself is available as Open Source.

## II. SYSTEM ARCHITECTURE

Fig. 1 shows the overall system architecture model of the proposed framework used to realize context-aware anticipatory communication mechanisms. Instead of sending data in a periodical way, a probability for the transmission of the whole data buffer is computed within each time step based on the knowledge about the context characteristics. The schemes operates at the application level and can therefore easily integrating into existing sensor applications. The *channel context parameters* are defined by the Long Term Evolution (LTE) downlink indicators Reference Signal Received Power (RSRP), Reference Signal Received Quality (RSRQ), Signal-to-noise-ratio (SNR) and Channel Quality Indicator (CQI). Furthermore, the *mobility context parameters* consisting of the vehicle's velocity and direction as well as position information and map knowledge. All context parameters are recorded during the transmission phases and form the training data for the machine learning based prediction mechanism. In order to take the application requirements into account, the behavior is configured with the *system parameters* that describe the transmission metric Φ, which is either directly mapped to one of the downlink indicators or is derived by a combination of multiple indicators. In [2], we evaluated the data rate prediction accuracy with different machine learning models and used the prediction results themselves as a CAT metric. Fig. 2 shows key results for the achieved prediction accuracy as well as for the the machine learning based CAT in a real-world evaluation. Using the data rate prediction as a metric for context-aware communication, the resulting mean data rate has been improved by up to 164% on highway scenarios. The results further showed that no single network quality indicator provides enough information on its own for identifying a connectivity hotspot with a high grade of certainty. Consequently, the machine learning based metric (M5T decision tree) achieves the highest data rate gains as it considers the available information as a whole and sets into relation to the payload size.

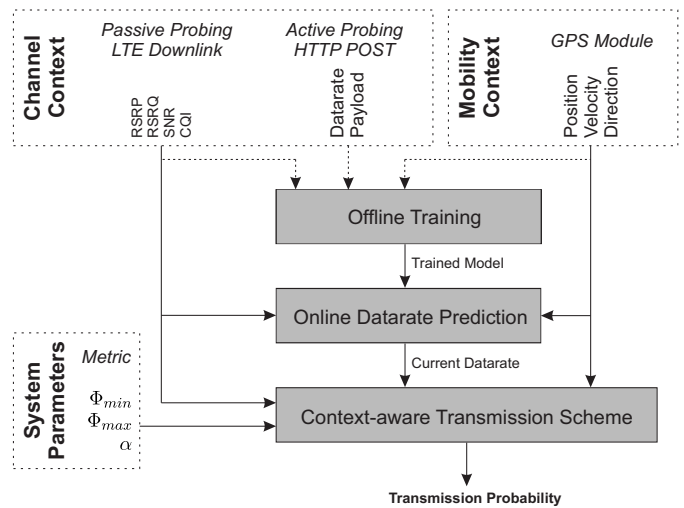

Fig. 1. Architecture model for the proposed context-aware communication framework. The abstract metric definition allows the application of the transmission scheme to different communication technologies and can be used to combine multiple transmission metrics.

## III. Demonstration

The demonstration at the conference venue will show the main features the context-aware transmission approach and will feature live capturing of network quality data as well as context-aware vehicular communication based on experimental traces.

### A. Data Acquisition and Live Analysis with an Android-based Measuring App

The first part of the demo will address the topic of obtaining network quality data in real world cellular networks. For this purpose, an Android-based measuring application will be presented that is able to capture and visualize the behavior of the current network quality indicators as well as performing and evaluating data transmissions with the CAT communication scheme. The application is freely available as Open Source[1] and is intended to be used to capture the required training data for the machine-learning based data rate prediction. Additionally, it serves as a starting point for the evaluation of novel context-aware transmission schemes.

[1]Available at https://github.com/BenSliwa/MTCApp

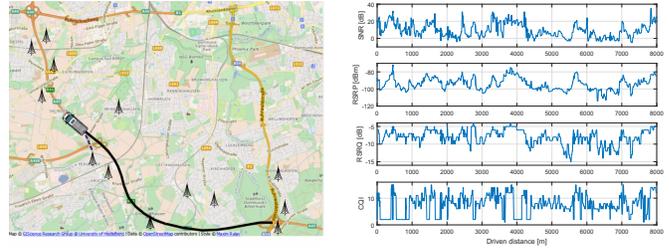

Fig. 3. Live-visualization of the mobility behavior and network quality indicators based on real-world trace data

### B. Visualization of the Real-world Context-aware Transmission Behavior in a Vehicular Scenario

The second part will show how the proposed resource-efficient transmission scheme behaves in a real-world vehicular context. For this purpose, we will utilize the obtained traces from the large-scale field evaluation and visualize the dynamic mobility behavior of the measurement vehicle on its driven trajectory as well as the channel conditions the vehicle is experiencing at the respective locations. The dataset consists of Key Performance Indicator (KPI) measurements from 4000 transmissions that were performed in a public cellular network with more than 1000 km total driven distance on is publicly available at [4].

## Acknowledgment

Part of the work on this paper has been supported by Deutsche Forschungsgemeinschaft (DFG) within the Collaborative Research Center SFB 876 "Providing Information by Resource-Constrained Analysis", projects A4 and B4 and has been conducted within the AutoMat (Automotive Big Data Marketplace for Innovative Cross-sectorial Vehicle Data Services) project, which received funding from the European Union's Horizon 2020 (H2020) research and innovation programme under the Grant Agreement No 644657. Thomas Liebig received funding from the European Union Horizon 2020 Programme (Horizon2020/2014-2020), under grant agreement number 688380 "VaVeL: Variety, Veracity, VaLue: Handling the Multiplicity of Urban Sensors".

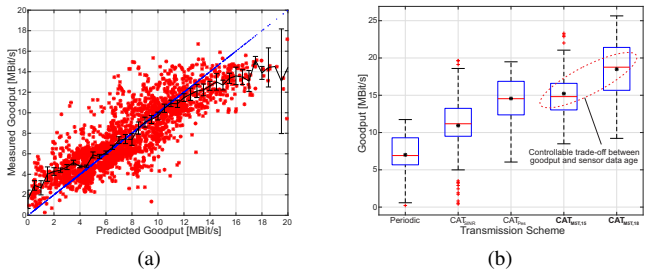

Fig. 2. Accuracy of the machine learning based data rate prediction and resource-efficiency improvements by leveraging the prediction as a metric for context-aware communication *Source: Adapted from [2]*